\documentclass[preprint]{aastex}

\usepackage{amsmath,amssymb,graphicx}

\shorttitle{Microlensing with advanced contour integration
algorithm}

\begin{document}

\title{Microlensing with advanced contour integration algorithm:
Green's theorem to third order, error control, optimal sampling
and limb darkening}

\author{V. Bozza\altaffilmark{1}}
\affil{Dipartimento di Fisica ``E. R. Caianiello", Universit\`a di
Salerno, I-84084 Fisciano, Italy. \\
 Istituto Nazionale di Fisica Nucleare, Sezione di Napoli,
 Italy. \\
 Istituto Internazionale per gli Alti Studi Scientifici, I-84019,
 Vietri sul Mare, Italy.}

 \altaffiltext{1}{valboz@sa.infn.it}

\begin{abstract}

Microlensing light curves are typically computed either by
ray-shooting maps or by contour integration via Green's theorem.
We present an improved version of the second method that includes
a parabolic correction in Green's line integral. In addition, we
present an accurate analytical estimate of the residual errors,
which allows the implementation of an optimal strategy for the
contour sampling. Finally, we give a prescription for dealing with
limb-darkened sources reaching arbitrary accuracy. These
optimizations lead to a substantial speed-up of contour
integration codes along with a full mastery of the errors.
\end{abstract}

\keywords{gravitational lensing –- methods: numerical –- binaries:
general –- planetary systems}

\section{Introduction}

Microlensing is one of the most promising methods for finding the
first Earth-like extrasolar planet \citep{Gaudi2010,DomRev}. When
a compact object transits very close to the line of sight of a
background source star, the flux coming from the source is
amplified by gravitational lensing and follows a typical
bell-shape light curve, analytically described by
\citet{Pacz1986}. If the lens is a star accompanied by a secondary
body like a planet, an additional bump or dip appears on the light
curve \citep{MaoPacz}. The timescale of such features ranges from
a few hours to a few days, depending on the square root of the
mass of the planet \citep{GouLoe}. Such short timescales require
intensive monitoring by telescopes situated all over the Earth.

Nowadays, more than 30 telescopes are involved in microlensing
searches towards the Galactic bulge. More than 600 events are
discovered every year. Out of these, roughly from 10 to 20 events
show anomalies that can be interpreted as due to binary lenses.
Some of these are finally accepted as showing evidence of an
extrasolar planetary system. Since the start of microlensing
searches, 26 events have been reported as containing planetary
candidates with stronger or weaker evidence \cite{DomRev}. Nine of
these events have also been included in the most updated exoplanet
list available on the web (http://exoplanet.eu)
\citep{Beaulieu2006,Bennett2008,Bond2004,Dong2009,Gaudi2008,Gou2006,Janczak2009,Sumi2010,Udalski2005}.
Finally, one event brings the spectacular signature of two planets
in the same system \citep{Gaudi2008}.

Unfortunately, interpreting binary microlensing events is a very
long process that may even take from one to four years for a
single event. This is due to several reasons: one is related to
the difficulty of getting rid of all systematic errors in the
photometry of each dataset. Some sets of images need to be reduced
with different methods in order to compare the effects of
systematics, and subtract them from the final result. After the
reduction process is complete, the final datasets may contain
hundreds or thousands of data points, which are then ready for the
modelling process.

The modelling process is typically driven by the criterium of
$\chi^2$ minimization, which can be achieved either by downhill
algorithms or by Markov Chain Monte Carlo (MCMC) methods. In order
to evaluate the $\chi^2$ for a single tentative model, it is
necessary to compute the microlensing magnification at each data
point. However, as is well known, microlensing magnification
cannot be calculated analytically if the lens is a binary system.
As the angular extension of the source plays a major role in the
observed magnification, the computation of even one single model
point is relatively time-consuming. Multiplying this basic time
unit for the number of data points, the number of models within a
MCMC simulation, the number of different hypotheses to be checked
(parallax, xallarap, orbital motion, limb darkening, binary
source, $\ldots$), we can easily imagine why the study of a single
event takes so long. It is then critical to reduce the
computational time of a single model point as much as possible, so
that the whole modelling process is cut down to a more reasonable
duration.

There are basically two classes of methods for the computation of
the microlensed flux of a source. The first is based on the
construction of ray-shooting maps \citep{Kayser}. In practice,
rays are shot back from the observer to the lens plane and then
deflected to the source plane. If they intercept the source disk
they are counted as contributing to the total magnification. This
method has three main advantages: it is conceptually simple, it
can naturally take into account the limb darkening profile of the
source, maps at fixed lens configurations can be re-used for
different source positions. Numerous optimized versions have
appeared in the literature, designed for re-use of maps
\citep{Wambsganss,Rattenbury,Dong2006}. An alternative strategy is
to speed-up the computation for a single source position giving up
the map re-use \citep{BenRhi}. In this case, starting from the
positions of the centers of the images, one shoots rays only where
really needed. A further improvement of this method employing a
polar coordinate-grid with an optimized prescription to handle
limb darkening has been recently presented by \citet{Bennett2009}.

The second method is based on an application of Green's theorem
(which can be viewed as a two-dimensional version of Stokes'
theorem). In practice, one can find the area of an image by
calculating a Riemann integral along the image contour. The use of
contours in gravitational lensing dates back to \citet{SchKay}.
Green's theorem was then used for calculating areas by
\citet{Dom93}. The method has been refined by \citet{Dom95} and
then applied to microlensing by \citet{GouGau} and \citet{Dom98}.
The appeal of this method is that a two-dimensional calculation is
turned into a one-dimensional calculation, which is much faster,
in principle. Related to this approach are the algorithms
presented by \citet{Dong2006} and the adaptive grid search by
\citet{Dominik2007}. The main advantages come from the potentially
high computational velocity and the high flexibility for models in
which the lens configuration changes (e.g. in the treatment of
planet orbital motion). However, the contour integration approach
requires an images reconstruction procedure (which can be
sometimes complicated); in addition, limb darkening cannot be
naturally incorporated in the algorithm. In particular, the latter
limitation has oriented the community to give a general preference
to ray-shooting methods.

Nevertheless, apart from its undisputed elegance, the contour
integration approach is still competitive for obtaining
preliminary microlensing models very quickly, which is
particularly interesting in view of the realization of real-time
modelling of binary microlensing events. Furthermore, when orbital
motion is relevant, traditional ray-shooting methods typically
become definitely too heavy. In this case, only adaptive methods
\citep{BenRhi,Bennett2009} can compete with Green's theorem
algorithms.

Finally, besides ray-shooting and contour integration methods, it
is worth mentioning that when the source size is only marginally
relevant (for sources not too close to caustics), one can
approximate its effects by quadrupole or hexadecapole
approximations \citep{Gould2008,PejHey}. These methods allow to
obtain a substantial speed-up of the code avoiding useless heavy
computations when the source size correction is small. They can be
used in combination with other methods that may intervene when the
source gets closer to a caustic.

In this paper we present four new ideas for boosting codes based
on contour integration approach. In Section 2 we show how Green's
line integral can be approximated to third order introducing a
parabolic correction, with a substantial improvement in accuracy.
In Section 3 we present accurate estimates of the residual errors
in Green's integral. These are used to implement an optimal
sampling strategy that allows to minimize the calculations for a
given required accuracy, as explained in Section 4. In Section 5
we suggest an easy prescription for the treatment of limb
darkening that achieves a fixed accuracy avoiding lengthy
calculations. The benefits achieved by all these innovations are
documented by several numerical examples in Section 6.

\section{Green's line integral to third order}

\subsection{The concept of Green's line integral}

Consider a generic continuous gravitational lens mapping between
the image plane $\vec x$ and the source plane $\vec y$
\begin{equation}
\vec y = \vec f (\vec x ). \label{Lens Map}
\end{equation}

Consider a circular source $A_S$ with radius $\rho_*$ centered in
the position $\vec y_S$. The boundary of the source is a circle of
radius $\rho_*$ that we shall indicate by $\gamma_S$. A trivial
parametrization of this curve is
\begin{equation}
\vec y(\theta)= \vec y_S+ \rho_* \left(\begin{array}{c}
  \cos \theta \\
  \sin \theta \\
\end{array} \right).
\end{equation}

For each $\theta$, we can solve the lens equation (\ref{Lens Map})
with $\vec y =\vec y(\theta)$. As $\theta$ runs from 0 to $2\pi$,
the solutions of this equation describe several curves $\gamma_I$
in the image plane, parameterized as $\vec x_I(\theta)$. The
subscript $I$ runs from 1 to the number of images $N$. In the case
of a binary lens, $N=3$ if the source is outside all caustics,
$N=5$ if the source is completely inside a caustic. If part of the
source is inside a caustic, then two images are created at some
$\theta_c$ and disappear at some $\theta_d$, so that $N=5$ with
two images $\vec x_I(\theta)$ defined only in the subinterval
$[\theta_c,\theta_d]$. Creation-destruction of images may also
occur in several disjoint subintervals of $[0,2\pi]$, if the
source touches two or more caustics. All curves $\gamma_I$ have
definite parity $p_I=\pm 1$ and represent the boundaries of the
regions in the image plane that are mapped to the source $A_S$
through the lens map $\vec f$. Such regions represent the physical
images of our source. The ratio between the total area $A$ of all
images and $A_S$ represents the sought magnification factor.

By Green's theorem, the area enclosed by a closed curve $\gamma$
is
\begin{equation}
A = \pm\frac{1}{2}\int\limits_{\gamma} \vec x \wedge d \vec x,
\end{equation}
where the positive sign is taken for counterclockwise curves and
the negative sign is taken for clockwise curves. We remind that
the wedge product between two vectors is a pseudoscalar in two
dimensions: $\vec x \wedge \vec y \equiv x_1 y_2-x_2 y_1$.

As our parametrization of the source boundary $\gamma_S$ is
counterclockwise, positive parity $\gamma_I$'s are still
counterclockwise, whereas negative parity $\gamma_I$'s are
clockwise. Therefore, the total area of all images can be found as
\begin{equation}
A = \sum\limits_I \frac{1}{2}p_I\int\limits_{\gamma_I} \vec x_I
\wedge d \vec x_I. \label{Stokes}
\end{equation}

Such expression still holds also when part of the source is inside
a caustic \citep{Dom95,GouGau}.

\subsection{Trapezium approximation of Green's integral}

In order to find a numerical approximation to Eq.  (\ref{Stokes}),
we must introduce a sampling of the source boundary in the
following way
\begin{equation}
\vec y_i= \vec y_S+ \rho_* \left(\begin{array}{c}
  \cos \theta_i \\
  \sin \theta_i \\
\end{array} \right),
\end{equation}
where $\left\{\theta_i\right\}$ is an arbitrary ordered sequence
of $n$ numbers with $0=\theta_0<\theta_1< \ldots < \theta_i<
\ldots < \theta_n=2\pi$. One simple possibility is to take a
uniform sampling $\theta_{i+1}-\theta_i=\mathrm{const}$. However,
this is not necessary and more optimal choices are possible, as
will be explained in Section \ref{Sec Sampling}.

For each $\theta_i$, we solve the lens equation (\ref{Lens Map})
and find the corresponding points $\vec x_{I,i}$ on the image
boundaries $\gamma_I$. If $\vec x_{I,i+1}$ is close enough to
$\vec x_{I,i}$, it makes sense to approximate Eq. (\ref{Stokes})
as
\begin{eqnarray}
&A & \simeq \frac{1}{4} \sum\limits_I p_I \sum\limits_{i=0}^{n-1}
\left(\vec x_{I,i+1}+\vec x_{I,i} \right) \wedge \left(\vec
x_{I,i+1}-\vec x_{I,i} \right) = \nonumber \\ && \frac{1}{2}
\sum\limits_I p_I \sum\limits_{i=0}^{n-1} \vec x_{I,i} \wedge \vec
x_{I,i+1}= \nonumber \\&& \frac{1}{2} \sum\limits_I p_I
\sum\limits_{i=0}^{n-1} \left( x_{I,i+1,2} + x_{I,i,2}\right)
\left(x_{I,i,1} - x_{I,i+1,1} \right), \label{Stokes 1st}
\end{eqnarray}
where the last version is simply the trapezium approximation of
the Riemann integral of the function $x_{I,2}(x_{I,1})$. It is
more advantageous numerically in that it has one multiplication
instead of two.

Eq. (\ref{Stokes 1st}) is written in the case of no caustic
crossing. It can be easily extended to the general case by letting
$i$ run only on the values for which the image $\gamma_I$ exists
and adding up connection terms between each pair of created images
and each pair of destroyed images (see also Section \ref{Sec Parab
crit}).

Summing up, in the implementation of the trapezium approximation
of Green's line integral (\ref{Stokes 1st}), we need the following
routines:
\begin{itemize}
\item{A routine solving the lens equation for each source position
$\vec y_i$. For example, one can use the zroots routine of
Numerical Recipes \citep{NumRec}.}%
\item{A routine associating the solutions $\vec x_{J,i+1}$ found
at each $(i+1)-$th step with the correct image $\gamma_I$. This
can be done by re-ordering the solutions in such a way that $|\vec
x_{I,i}-\vec x_{I,i+1}|<|\vec x_{I,i}-\vec x_{J,i+1}|$ for each
$J\neq I$. If new images are created or destroyed, they will be
recognized as the last two unmatched images. Of course, this
association routine has some failure probability when the new
solutions $\vec x_{J,i+1}$ are too far from the old ones $\vec
x_{I,i}$. However, we will see in Section \ref{Sec Errors} that a
careful estimation of the errors will easily recognize such
situations.
}%
\end{itemize}

We find that roughly $80\%$ of the machine time is spent in the
root finding routine, for which there is basically no hope of
further optimization (we already re-use old roots as starting
values for the next calculation). So, the only way to speed up a
contour integration code is to reduce the number of points in the
sampling while keeping the same accuracy. This can be achieved by
pushing the numerical approximation of Green's integral to higher
orders.

Another possibility to get around the problem of root finding is
the use of adaptive grids on the lens plane \cite{Dominik2007}. In
these algorithms, the sampling of the image boundaries is obtained
by a grid construction directly on the lens plane. Although we
will mostly refer to the scheme described in this subsection
(source sampling and lens equation solving to obtain an image
sampling), most of the concepts introduced in this paper can also
be applied to algorithms based on direct sampling on the lens
plane.

\subsection{Parabolic correction of Green's integral}

Going back to Eq. (\ref{Stokes}), we can write it as
\begin{equation}
A = \sum\limits_I \frac{1}{2}p_I\int\limits_{0}^{2\pi} \vec x_I
\wedge \vec x'_I d\theta,
\end{equation}
where the prime denotes derivation with respect to the parameter
$\theta$.

Let us consider the generic image $\gamma_I$ and the generic
interval $[\theta_i,\theta_{i+1}]$, with size $\Delta\theta$. The
contribution of this interval to the whole integral is
\begin{equation}
dA_I = \frac{1}{2}\int\limits_{\theta_i}^{\theta_i+\Delta\theta}
\vec x_I \wedge \vec x'_I d\theta. \label{dAI}
\end{equation}

The first order trapezium approximation used up to now just reads
(see Eq. (\ref{Stokes 1st}))
\begin{equation}
dA_I^{(t)} = \frac{1}{2} \vec x_I(\theta_i) \wedge \vec
x_I(\theta_i+\Delta\theta). \label{dAIt}
\end{equation}
Comparing the expansions of $dA_I$ and $dA_I^{(t)}$ in powers of
$\Delta\theta$, we find that they coincide at the first and second
order, the difference being of order $\Delta\theta^3$.

Now, let us introduce the following correction term
\begin{equation}
dA_I^{(p)} = \frac{1}{24}\left[\left.\left(\vec x'_I\wedge \vec
x''_I\right)\right|_{\theta_i} +\left.\left(\vec x'_I\wedge \vec
x''_I\right)\right|_{\theta_{i}+\Delta
\theta}\right]\Delta\theta^3. \label{dAIp}
\end{equation}
Adding this correction to the trapezium approximation and
comparing the power expansion to that of the exact integral
(\ref{dAI}), we have
\begin{equation}
dA_I=dA_I^{(t)}+dA_I^{(p)} +O(\Delta\theta^5).
\end{equation}

The residual error is now of order $\Delta\theta^5$, which is much
smaller than what can be achieved by the trapezium approximation.
$dA_I^{(p)}$ can be viewed as a parabolic correction as it takes
into account the local curvature of $\gamma_I$ stored in the
second derivative $\vec x''$.

\subsection{Implementation of the parabolic correction}

$dA_I^{(p)}$ is expressed in terms of derivatives with respect to
$\theta$ calculated at $\theta_i$ and $\theta_{i+1}$. In
principle, these derivatives can be easily calculated analytically
in terms of local quantities. The explanation is easier if we
switch to complex notations \citep{Witt90}. The coordinates in the
source and lens planes respectively become
\begin{eqnarray}
&& \zeta=y_1+i y_2 \\
&& z=x_1+i x_2.
\end{eqnarray}

The lens equation for a binary lens with mass ratio $q$ and
separation $a$ assumes the form
\begin{equation}
\zeta = z - \frac{1}{1+q}\left(\frac{1}{\bar z +
a/2}+\frac{q}{\bar z - a/2} \right), \label{LEQ complex}
\end{equation}
from which we obtain
\begin{eqnarray}
&& \partial\zeta/\partial z=1 \\
&& \frac{\partial\zeta}{\partial \bar z}=
\frac{1}{1+q}\left(\frac{1}{(\bar z +
a/2)^2}+\frac{q}{(\bar z - a/2)^2} \right)\\
&& \frac{\partial^2\zeta}{\partial \bar z^2}=-
\frac{2}{1+q}\left(\frac{1}{(\bar z + a/2)^3}+\frac{q}{(\bar z -
a/2)^3} \right).
\end{eqnarray}

The Jacobian determinant is
\begin{equation}
J= 1-\left|\frac{\partial\zeta}{\partial \bar z}\right|^2.
\end{equation}
Note that the Jacobian determinant must be calculated in the
linear approximation too, in order to assess the parity of the
image.

Deriving Eq. (\ref{LEQ complex}) with respect to $\theta$, we have
\begin{equation}
\zeta' = z' + \frac{\partial\zeta}{\partial \bar z} \bar z'.
\label{zeta' eq}
\end{equation}

Inverting this equation with its complex conjugate, we get
\begin{equation}
z' = \left[\zeta' - \frac{\partial\zeta}{\partial \bar z} \bar
\zeta'\right] J^{-1}.
\end{equation}

Deriving again with respect to $\theta$, we get the expression for
$z''$
\begin{equation}
z'' = \left\{\zeta'' - \frac{\partial^2\zeta}{\partial \bar z^2}
(\bar z')^2 - \frac{\partial\zeta}{\partial \bar z}\left[ \bar
\zeta'' - \frac{\partial^2 \bar \zeta}{\partial  z^2} ( z')^2
\right] \right\}J^{-1}.
\end{equation}

The key fact is that all these quantities can be calculated
exactly starting from the source parametrization, which in complex
notations reads
\begin{equation}
\zeta= \zeta_S+\rho_* e^{i \theta},
\end{equation}
where $\zeta_S=y_{S,1}+i y_{S,2}$ is the center of the source
disc. From this expression, we get $\zeta'= i \rho_* e^{i \theta}$
and $\zeta''= - \rho_* e^{i \theta}$.

The parabolic correction (\ref{dAIp}) contains terms of the type
\begin{equation}
\vec x'\wedge \vec x'' = \frac{1}{2i}\left( z''\bar z' - z'\bar
z'' \right). \label{z'z''}
\end{equation}
Plugging the former expressions for $z'$ and $z''$ and using the
source parametrization, we finally obtain the compact expression
\begin{equation}
\vec x'\wedge \vec x'' = \left\{ \rho_*^2 + \mathrm{Im}\left[
(z')^2 \zeta' \frac{\partial^2 \bar \zeta}{\partial z^2} \right]
\right\} J^{-1}. \label{x'x''}
\end{equation}

The implementation of a parabolic correction is therefore
relatively simple. For each $\theta$, after the extraction of the
roots of the lens equation, we just have to calculate $z'$, $J$
and $\partial^2 \bar \zeta/\partial z^2$ for each root, taking
$\zeta'=i \rho_* e^{i\theta}$, and then store the value of the
wedge product (\ref{x'x''}).

Finally, when we compute Green's line integral, for each arc
$[\vec x_{I,i},\vec x_{I,i+1}]$ we can put together all the
ingredients to calculate both the trapezium approximation
(\ref{dAIt}) and the parabolic correction (\ref{dAIp}).

\subsection{Parabolic correction at critical points} \label{Sec
Parab crit}

As pointed out before, it might happen that a portion of the
source boundary lies inside a caustic. In this case, at some
$\theta_c$ a pair of new images is created and at $\theta_d$
another pair is destroyed. Green's theorem can still be applied,
but since $\theta_c$ and $\theta_d$ do not generally belong to our
sampling $\{\theta_i\}$, we need to introduce connection terms
between the starting points of the created images (the same
happens for the pair of destroyed images). Let us discuss the case
for pair creation, the destruction being analogue.

For a given sampling $\{\theta_i\}$, the new pair of images
appears at some $\theta_i$, with $\theta_{i-1}<\theta_c<\theta_i$.
Let us call the starting points of the new images $\vec x_{+,i}$
and $\vec x_{-,i}$. The problem is that the parametric distance
between the two images is not available, since the precise value
of $\theta_c$ is unknown. However, the standard expansion of the
lens equation in a neighborhood of a fold tells us that the two
created images move away from the creation point as \citep{SEF}
\begin{equation}
\vec x_\pm=\left( \begin{array}{c}
  \frac{y_1}{\lambda} \\
  \\
  \frac{-b y_1 \pm\sqrt{2ay_2
\lambda^2 +(b^2-ac)y_1^2}}{a\lambda} \\
\end{array}\right), \label{xfold}
\end{equation}
where $\lambda$, $a$, $b$ and $c$ are coefficients of the
expansion of the lens equation near a fold, $(y_1,y_2)$ is the
position of the source relative to the fold caustic and $\vec
x_\pm$ is the position of each of the two images relative to the
critical curve.

Expanding our parametrization in the neighborhood of the crossing
point, we have $\vec y= (\theta-\theta_c) (c_1,c_2)$, with $c_1$
and $c_2$ being two constants depending on $\rho_*$ and
$\theta_c$. The sought connection term reads
\begin{equation}
dA_c= \int\limits_{\theta_c}^{\theta_i} \vec x_+ \wedge \vec x'_+
dt - \int\limits_{\theta_c}^{\theta_i} \vec x_- \wedge \vec x'_-
dt,
\end{equation}
where we have assumed that $\vec x_+$ is the positive parity
solution while $\vec x_-$ is the negative parity one.

The trapezium approximation with the correct signs is simply
\begin{equation}
dA_c^{(t)} = \frac{1}{2} \left( x_{-,i,2} + x_{+,i,2}\right)
\left(x_{-,i,1} - x_{+,i,1} \right).
\end{equation}

Now, we propose the parabolic correction
\begin{equation}
dA_c^{(p)} = \frac{1}{24}\left[\left(\vec x'_{+,i}\wedge \vec
x''_{+,i}\right) -\left(\vec x'_{-,i}\wedge \vec x''_{-,i}\right)
\right] \widetilde{\Delta\theta}^3, \label{dAIpc}
\end{equation}
where
\begin{equation}
\widetilde{\Delta\theta} =  \frac{\left| \vec x_{+,i} - \vec
x_{-,i} \right| }{\sqrt {|\vec x'_{+,i} \cdot \vec x'_{-,i}|}}
\label{tildedtheta}
\end{equation}
replaces the parametric distance $\Delta\theta$ used in the
ordinary parabolic correction.

Using the approximate general expressions for the images
(\ref{xfold}), and expanding $dA_c$, $dA_c^{(t)}$ and $dA_c^{(p)}$
in powers of $(\theta_i-\theta_c)$, we realize that the trapezium
approximation is accurate only to first order in
$(\theta_i-\theta_c)$, the residual error being of order
$(\theta_i-\theta_c)^{3/2}$. The parabolic correction accounts for
the term of order $(\theta_i-\theta_c)^{3/2}$ and leaves a
residual error of order $(\theta_i-\theta_c)^{5/2}$. Note that the
orders of the errors in arcs containing a pair creation or
destruction are halved with respect to ordinary arcs. This is a
major reason for increasing the sampling of the source boundary
near caustic points. With a uniform sampling, instead, the error
would be largely dominated by intervals containing caustic
crossings.
%
%As a final remark, we note that when the fold crossing falls very
%close to a cusp
%
%
%source contains a cusp, its boundary intercepts the two folds
%adjacent to the cusp and our estimate of the errors is always
%valid.

\subsection{Final remarks on the parabolic correction}

The parabolic correction has some computational cost because it
requires some additional operations to be performed on each root.
However, such cost remains negligible with respect to the time
spent in the root inversion routine. Moreover, it helps reducing
the residual error dramatically for each sampling interval
$[\theta_{i},\theta_{i+1}]$ up to the fifth order in
$\Delta\theta$. We can read this achievement in the other way
round: with the parabolic correction we can reach the same
accuracy as with the trapezium approximation only, but with a much
sparse sampling of the boundary curves. Since, as said before,
most of the computational time is spent in the root inversion
routine, which must be run once for each $\theta_i$, we can aim at
a substantial speed up of the code if we are able to reduce the
sampling as much as we can without losing accuracy thanks to the
parabolic correction. A fundamental step toward this goal is a
careful estimation of the residual errors, which is the subject of
the next section.

\section{Error Estimators} \label{Sec Errors}

\subsection{Error estimators for ordinary images}

As shown in the previous section, the residual errors in Green's
line integral after the introduction of the parabolic correction
are of the order $\Delta\theta^5$. The exact expression for the
fifth order term in the power expansion of Green's integral
(\ref{dAI}) contains third derivatives of $z$, as can be easily
guessed. However, we do not want to make more calculations for the
estimate of the fifth order term and rather use the quantities
already calculated to make a realistic but economic estimate.
Secondly, higher and higher order derivatives are more and more
affected by numerical errors. Finally, we must also take into
account the possibility of a wrong matching of the images, as
anticipated in the previous section.

For all these reasons, we disregard the fifth order term in
Green's integral and prefer to introduce three new quantities as
error estimators. These three quantities are built up from first
and second derivatives of $z$, thus requiring a minimum amount of
additional calculations. They are meant to intervene in different
situations with the aim of being complementary to each other and
cover all possible sources of error.

The first estimator is
\begin{equation}
E_{I,i,1} = \frac{1}{48}\left|\left.\left(\vec x'_I\wedge \vec
x''_I\right)\right|_{\theta_i} -\left.\left(\vec x'_I\wedge \vec
x''_I\right)\right|_{\theta_{i+1}}\right|\Delta\theta^3.
\label{E1}
\end{equation}

As the parabolic correction is based on an average of the wedge
product $\vec x'_I\wedge \vec x''_I$ on the two end points of the
arc, it is natural to estimate the error using the difference of
the two quantities that are averaged. A power expansion in
$\Delta\theta$ reveals that $E_1$ is of order $\Delta\theta^4$
rather than $\Delta\theta^5$. The main reason for using
$E_{I,i,1}$ is that it performs very well in the identification of
wrong images matching, since the two wedge products take very
different values if the two points do not belong to the same
image. On the other hand, it does not seem to systematically
dominate over fifth order error estimators.

The second estimator is
\begin{equation}
E_{I,i,2} = \frac{3}{2}\left| dA_I^{(p)}\left(\frac{\left|\vec
x_{I,i}-\vec x_{I,i+1}\right|^2}{\Delta\theta^2 \left| \vec
x'_{I,i}\cdot \vec x'_{I,i+1} \right|} -1\right) \right|
\label{E2}.
\end{equation}
The ratio of the squared distance between the two points and the
scalar product of their derivatives is an approximation to
$\Delta\theta^2$ (see the definition of $\widetilde{\Delta\theta}$
in equation (\ref{tildedtheta})). It can be shown that $E_{I,i,2}$
is of order $\Delta\theta^5$. It also takes large values in case
of wrong images matching but works in a complementary way to $E_1$
as it is based on different quantities. In particular, this
estimator is particularly effective for the detection of hidden
cusp crossings between $\vec y_i$ and $\vec y_{i+1}$, which might
otherwise be a very dangerous situation.

The last estimator is
\begin{equation}
E_{I,i,3} = \frac{1}{10}\left| dA_I^{(p)}\right |\Delta\theta^2
\label{E3},
\end{equation}
which intervenes in undersampled situations when $E_{I,i,2}$ is
incidentally zero.

The error estimators just defined can be combined so as to build
an error estimate for each arc $[x_{I,i},x_{I,i+1}]$. We adopt a
simple sum
\begin{equation}
E_{I,i}=E_{I,i,1}+E_{I,i,2}+E_{I,i,3}.
\end{equation}
We prefer a simple sum to the usual quadrature combination of the
errors in order to minimize the operations while keeping a more
conservative attitude.

\subsection{Error estimators at critical points}

For the connection terms between image pairs created at some
critical points we need different error estimators.

We define
\begin{equation}
E_1^{(c)} = \frac{1}{48}\left|\left(\vec x'_{+,i}\wedge \vec
x''_{+,i}\right) +\left(\vec x'_{-,i}\wedge \vec x''_{-,i}\right)
\right| \widetilde{\Delta\theta}^3, \label{E1c}
\end{equation}
in analogy to $E_{I,i,1}$. This quantity is of order 2 in
$(\theta_i-\theta_c)$, instead of order $5/2$. The same comments
as for $E_1$ apply.

The second estimator is
\begin{eqnarray}
&E_2^{(c)} &= \frac{3}{2}\left| \left( \vec x_{+,i} - \vec x_{-,i}
\right)\cdot \left(\vec x'_{+,i} - \vec x'_{-,i}\right) \right.
\nonumber \\ && \left. \mp \left| \vec x_{+,i} - \vec x_{-,i}
\right|\sqrt{|\vec x'_{+,i} \cdot \vec x'_{-,i}|}\right|
\widetilde{\Delta\theta} \label{E2c},
\end{eqnarray}
which is still of second order in $(\theta_i-\theta_c)$. The upper
sign applies at creation of two images and the lower sign applies
for destruction. Indeed, at the creation of two images the two
starting points are expected to move far apart in opposite
directions. Conversely, at destruction of two images, the end
points converge to the same critical point. $E_2^{(c)}$ becomes
very large if this is not the case.

The third estimator is
\begin{equation}
E_3^{(c)} = \frac{1}{10}\left| dA_c^{(p)}\right|
\widetilde{\Delta\theta}^2,
\end{equation}
which is analogous to $E_{I,i,3}$ and is of order $5/2$ in
$(\theta_i-\theta_c)$.

\subsection{Resurrecting buried pairs of images} \label{Sec ghost}

Finally, a very dangerous situation may occur when working with
Green's theorem approach and a very sparse sampling, which might
be likely if we take full advantage of the parabolic correction.
If the source boundary grazes a caustic, it might happen that a
very small slice of the source is inside it. This slice might
entirely fall in the middle between two sampling points, so that
$\theta_i<\theta_c<\theta_d<\theta_{i+1}$. In this situation, we
have no idea that the $i$-th interval contains an additional pair
of images. These images could be completely missed with a
consequent dangerously large error, whereas all remaining images
are calculated to the desired precision.

In order to thwart this menace, we introduce an additional
estimator in the following way. Suppose that at $\theta_i$ the
source boundary is outside the caustic. Then we have three real
images satisfying the complex lens equation plus two additional
roots that satisfy the fifth order polynomial version of the lens
equation but not the original one. When the source crosses a
caustic, these two ghost roots merge into a double root and then
become real.

Therefore, we can estimate how far $\vec y_i$ is from a caustic
evaluating the distance between these two ghost roots
\begin{equation}
g_i=|z_{g1,i}-z_{g2,i}|.
\end{equation}
Our idea is to check whether $g_{i-1}-g_i>g_i$. By linear
interpolation one would expect that at step $i+1$ the source is
inside the caustic and the two ghost roots have become real. If
this is not the case, we add an error
\begin{equation}
E_{G,i}=(g_{i-1}-g_i)^2,
\end{equation}
which might be quite large. As we shall see in the next section,
this large error will drive the optimal sampling strategy to
oversample the $i$-th interval in search for a possible caustic
crossing.

Of course, for symmetry, we also check that $g_{i+1}-g_i>g_i$ and
add an error $E_{G,i-1}=(g_{i+1}-g_i)^2$ if the source is outside
the caustic at step $i-1$. This check on the ghost images turns
out to be very effective in discovering buried caustic crossings
in undersampled regions.

\section{Optimal sampling} \label{Sec Sampling}

A uniform sampling of the images amounts to adopting a fixed step
size $\Delta\theta$, so that $\theta_{i+1}-\theta_i=\Delta\theta$
for each $i$. A uniform sampling may be quite inefficient: in
fact, it might oversample regions with little contribution to the
magnification or with small errors whereas caustic crossings,
which require a denser sampling, would not receive any particular
regard.

For these reasons, we adopt a different sampling strategy.
Consider a given $n$-points sampling $\{\theta_1, \ldots,
\theta_n\}$, with $\theta_1=0$ and $\theta_n=2\pi$. The question
is where to put the next sampling point $\hat \theta$.

For each interval $[\theta_i,\theta_{i+1}]$, an estimate of the
errors is given by
\begin{equation}
E_{i}=\sum\limits_I E_{I,i},
\end{equation}
where $I$ runs on all images present in the $i$-th interval and
the error estimators $E_{I,i}$ are defined in Section \ref{Sec
Errors}. If any images are created or destroyed in this interval,
we also add the errors of the corresponding connection term.
Moreover, we also perform the additional check described in
Section \ref{Sec ghost} and add the corresponding error term if
the source is grazing a caustic.

Then, we select the interval (labelled by $\hat i$) with the
largest error ($E_{\hat i}> E_{i}$ for all $i\neq \hat i$).
Finally, we simply set the new sample point as the midpoint of the
interval with the largest error: $\hat\theta=(\theta_{\hat
i}+\theta_{\hat i+1})/2$.

The sampling is thus increased only where really needed, starting
with the intervals with the largest errors. Since the new sample
point lies in the middle of the sampling sequence, once we
calculate the images $\vec x(\hat\theta)$ of $\vec y(\hat\theta)$,
we need to make the correct association both with the images $\vec
x(\theta_{\bar i})$ that precede $\vec x(\hat\theta)$ and with the
images $\vec x(\theta_{\bar i+1})$ following, so as to have the
new re-sampled boundary curves. After the association is done, we
can recalculate the magnification contributions and the errors in
the new sub-intervals $[\theta_{\hat i},\hat \theta]$ and $[\hat
\theta,\theta_{\hat i+1}]$. As can be easily guessed, this
procedure is technically easier to achieve by defining the
boundary curves and the sampling sequence as linked lists rather
than arrays, so that we can easily cut and link them so as to
insert new members in the middle.

As a starting minimal sequence, we take $\{0,\pi,2\pi\}$ with only
two points ($\theta=2\pi$ being just a replica of $\theta=0$). The
third sample point will thus be $\pi/2$ or $3\pi/2$ depending on
the respective errors of the two initial arcs, and so on.
Iterating our sampling procedure, we will end up with a
non-uniform sampling, with more points where the errors tend to
stay larger (typically close to caustic crossing points). If we
are far from any caustics, the sampling will tend to be uniform,
anyway.

The full control of the errors allows us to establish when to stop
the iteration. In fact, the total error in the magnification is
simply given by the sum of the errors of all intervals in the
sampling
\begin{equation}
E=\sum\limits_i E_{i}.
\end{equation}
Therefore, if our target accuracy is $\delta \mu$, we just have to
iterate until
\begin{equation}
E/(\pi \rho_*^2)<\delta\mu. \label{errcond}
\end{equation}

In some situations (at high magnification points), the images are
very thin stretched arcs, in which the errors of the inner side
compensate the errors of the outer side. In these cases, thanks to
this cancellation, the true error is much smaller than the sum of
the absolute values of the errors of all intervals. Our algorithm
can thus be stopped earlier than prescribed by Eq.
(\ref{errcond}). More precisely, we stop when the magnification
has changed by less than $\delta \mu/2$ in the last $n/2$ sampling
steps.

\section{Limb Darkening treatment} \label{Sec Limb}

The main drawback of codes based on contour integration approach
is that the source is treated as a uniform brightness disk. Real
stars are actually non-uniform, with a brightness profile that can
be approximated by a linear limb-darkening law \citep{Milne}
\begin{eqnarray}
&&I(\rho)=\bar I f(\rho/\rho_*) \nonumber \\
&&f(r)=\frac{1}{1-a/3}\left[1-a
\left(1-\sqrt{1-r^2}\right)\right],
\end{eqnarray}
where $r=\rho/\rho_*$ and $\bar I$ is the average surface
brightness, so that
\begin{equation}
\frac{1}{\pi \rho_*^2} \int\limits_0^{\rho_*} 2\pi \rho
I(\rho)d\rho = \bar I  \int\limits_0^{1} 2 r f(r)dr= \bar I.
\end{equation}
A typical profile is shown by the dotted line in Fig. \ref{Fig
Limb}.

\begin{figure}
\resizebox{10cm}{!}{\includegraphics{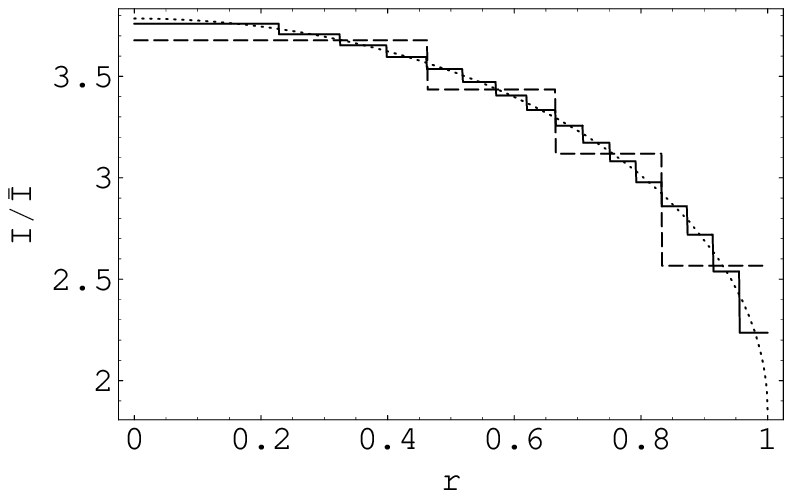}}
 \caption{Linear limb darkening profile with $a=0.51$ (dotted line)
 compared with a block approximation with four bins (dashed line) and 16 bins (solid line).}
 \label{Fig Limb}
\end{figure}

Indeed, whenever a caustic crossing is present in microlensing,
limb darkening cannot be neglected in accurate modelling of the
event. Therefore, if we want to use our Green's integral approach
with the parabolic correction in real microlensing events, we must
find a way to incorporate limb darkening. One obvious solution is
to calculate the magnification of several concentric disks at the
source position $\vec y_S$ with radii given by $\{\rho_1, \cdots,
\rho_m\}$, with $\rho_m=\rho_*$. Each disk should be weighted
according to the limb darkening profile in order to build up an
approximation to the correct result. In this scheme, for a single
point in the light curve, we must calculate the magnification of
$m$ disks instead of just one. As a consequence, the computation
is slowed down by a factor of $m$.

An alternative possibility, proposed by \citet{Dom98} is to
implement Green's theorem with different integrand functions
taking limb darkening into account. In any case, additional
integrations are required from the center of the source to the
periphery.

Unfortunately, there is no known way around this problem. Our
approach follows the standard multi-disk solution improved by a
clever choice of the radii $\rho_i$ and by a full control of the
errors.

\subsection{Magnification and errors in annuli}

Let us consider an annulus of our source with inner radius
$\rho_{i-1}$ and outer radius $\rho_i$. The total luminosity of
the source annulus is given by
\begin{equation}
 I_i^{(0)}=\int\limits_{\rho_{i-1}}^{\rho_i} 2\pi\rho  I(\rho)d\rho = \bar
 I\pi \rho_*^2
 [F(r_i)-F(r_{i-1})],
\end{equation}
where $r_i=\rho_i/\rho_*$ and we have introduced the cumulative
function
\begin{equation}
F(r)=2\int\limits_0^r dr' r'f(r').
\end{equation}

Gravitational lensing introduces a point-source magnification
factor $\mu(r,\theta)$ which modifies the observed luminosity as
\begin{equation}
 I_i=  \bar
 I \rho_*^2 \int\limits_{r_{i-1}}^{r_i} r f(r)
 dr\int\limits_0^{2\pi}
 d\theta\mu(r,\theta).
\end{equation}

The total observed luminosity is given by the sum of the
luminosities of all annuli. Dividing by the original source
luminosity $\bar I \pi \rho_*^2$, we get the limb-darkened
magnification factor
\begin{eqnarray}
&&M=\frac{1}{\bar I \pi \rho_*^2} \sum\limits_{i=1}^m I_i =\sum\limits_{i=1}^m M_i \\
&& M_i = \frac{1}{\pi}\int\limits_{r_{i-1}}^{r_i} r f(r)
 dr\int\limits_0^{2\pi}
 d\theta\mu(r,\theta). \label{Mi}
\end{eqnarray}

Using our contour integration approach, we are able to estimate
the magnification factor for a uniform disk of radius $\rho_i$ to
an arbitrary accuracy $\delta \mu$. Of course, such a
magnification factor for a finite size source is just the average
of the point-source magnification $\mu(r,\theta)$ on the source
disk
\begin{equation}
\mu_i=\frac{1}{\pi r_i^2}\int\limits_0^{r_i} r dr
\int\limits_0^{2\pi}
 d\theta\mu(r,\theta). \label{mui}
\end{equation}

Eq. (\ref{mui}) looks very similar to Eq. (\ref{Mi}), save for the
profile function $f(r)$ appearing inside the radial integration in
Eq. (\ref{Mi}). Therefore, we can approximate the contribution
$M_i$ of each annulus to the total magnification by replacing the
brightness profile $f(r)$ by a constant average brightness
\begin{equation}
f_i=\frac{F(r_i)-F(r_{i-1})}{r_{i}^2- r_{i-1}^2 }. \label{fi}
\end{equation}

Taking $f_i$ out of the integral, we get the following
approximation for $M_i$
\begin{equation}
\tilde M_i= f_i\left[\mu_i r_{i}^2- \mu_{i-1} r_{i-1}^2\right]
\end{equation}
in which all objects involved can be easily calculated in our
code. The approximate expression $\tilde M_i$ reduces to the exact
one $M_i$ for very thin annuli. As we can see in Fig. \ref{Fig
Limb}, the linear limb darkening profile is approximated by a
block function, in which each block has a constant brightness
given by $f_i$. Increasing the number of bins, the limb darkening
profile is approximated better and better. More specifically, if
$\delta r=r_i-r_{i-1}$, the difference between $\tilde M_i$ and
$M_i$ is of the third order
\begin{equation}
\delta M_i = \frac{r_i}{12\pi}f'(r_i)\delta r^3
\int\limits_0^{2\pi}d\theta
\partial_r\mu(r,\theta).
\end{equation}

This expression for the residual error can be used to construct an
efficient and economic error estimator without calculating
derivatives explicitly, namely
\begin{equation}
\delta \tilde
M_i^{(1)}=\left|\frac{1}{4}\left[r_i^2-r_{i-1}^2\right]\left[f(r_i)-f(r_{i-1})
\right] \left[\mu_i-\mu_{i-1} \right]\right|. \label{delta M1}
\end{equation}

Note that $\delta \tilde M_i^{(1)}$ reduces to $\delta M_i$ only
if we neglect the second and higher derivatives of $\mu(r,\theta)$
in a neighborhood of the source. This means that this error
estimator could be unreliable in some situations in which
$\mu(r,\theta)$ has a high curvature. We will come back to this
issue later.

At caustic crossings, $\mu(r,\theta)$ diverges and $\delta M_i$
loses meaning. In principle, our error estimator $\delta \tilde
M_i^{(1)}$ does not diverge but does not track the error
correctly. Therefore we introduce a new estimator that should be
used whenever the number of image contours at $r_{i-1}$ differs
from the number of contours at $r_i$
\begin{equation}
\delta \tilde
M_i^{(c)}=\left|\frac{1}{4}\left[r_i^2\mu_i-r_{i-1}^2\mu_{i-1}\right]\left[f(r_i)-f(r_{i-1})
\right] \right|, \label{delta Mc}
\end{equation}
which is always regular and of order $\delta r^2$ in the limit of
thin annuli.

\subsection{Sampling the source profile}

We have now an approximate form for the magnification of the
annuli and error estimators to control the accuracy. We must now
give a prescription for the choice of the radii of our annuli in
order to complete the limb darkening treatment.

Starting from a sequence ${r_1, \cdots, r_m}$, we select the
annulus with the largest error, say $[r_{\bar i -1},r_{\bar i}]$.
Then we divide it in two annuli, by inserting the radius $\bar r$
between $r_{\bar i -1}$ and $r_{\bar i}$. The new radius $\bar r$
is chosen in such a way that $F(r_{\bar i})-F(\bar r)=F(\bar
r)-F(r_{\bar i-1})$. In this way we make an equipartition of the
cumulative function. As a practical example of this partition
criterium of the source, in Fig. \ref{Fig Limb} we show a linear
limb darkening profile together with a block approximation with
four bins and a block approximation with 16 bins. The radii are
chosen so as to have $F(r_i)-F(r_{i-1})=1/n_{bins}$ and the
constant brightness value in each bin is given by Eq. (\ref{fi}).
We can see that this block approximation rapidly converges to the
exact profile when the number of bins is increased. Thanks to our
error control, however, we do not need to increase the sampling
everywhere but only where really needed. For example, if only the
periphery of the source intercepts a caustic, our error estimators
will require more annuli to be created close to $\rho_*$ without
calculating useless annuli at the center of the source.

Finally, let us come back to the issue of the second derivative of
$\mu(r,\theta)$. It might happen that the finite size
magnifications of the disks $\mu_i$ steadily grows from the center
to the periphery. However, it might also happen that at some
radius $r_{\bar i}$ $\mu_i$ starts to decrease. In such a
situation, it might happen that $\mu_{\bar i-1} \simeq \mu_{\bar
i}$ leading to a dangerously small $\delta \tilde M^{(1)}_{\bar
i}$. In order to overcome this problem, when adding a new radius
$\bar r$ to the sampling between $r_{\bar i-1}$ and $r_{\bar i}$,
with its finite size magnification $\bar \mu$, we calculate the
errors of the annuli $[r_{\bar i-1},\bar r]$ and $[\bar r,r_{\bar
i}]$ according to Eq. (\ref{delta M1}) or (\ref{delta M2}) and
then add to both annuli an error
\begin{equation}
\delta \tilde
M_i^{(2)}=\left|\frac{1}{4}\left[r_i^2-r_{i-1}^2\right]\left[f(r_i)-f(r_{i-1})
\right] \left[\mu_{\bar i}+\mu_{\bar i-1} -2 \bar
\mu\right]\right|, \label{delta M2}
\end{equation}
which accounts for possible changes of slope in the finite size
magnifications $\mu_i$.

Summing up, with this error-driven sampling strategy we continue
adding annuli until the total estimated error drops below the
desired accuracy $\delta \mu$. Each finite size magnification
$\mu_i$ is also calculated at accuracy $\delta \mu$. Since each of
them is weighted by the average flux in the expression of $\tilde
M_i$, the total error in $M$ coming from the $\mu_i$ is kept below
$\delta \mu$.

\begin{figure}
\resizebox{10cm}{!}{\includegraphics{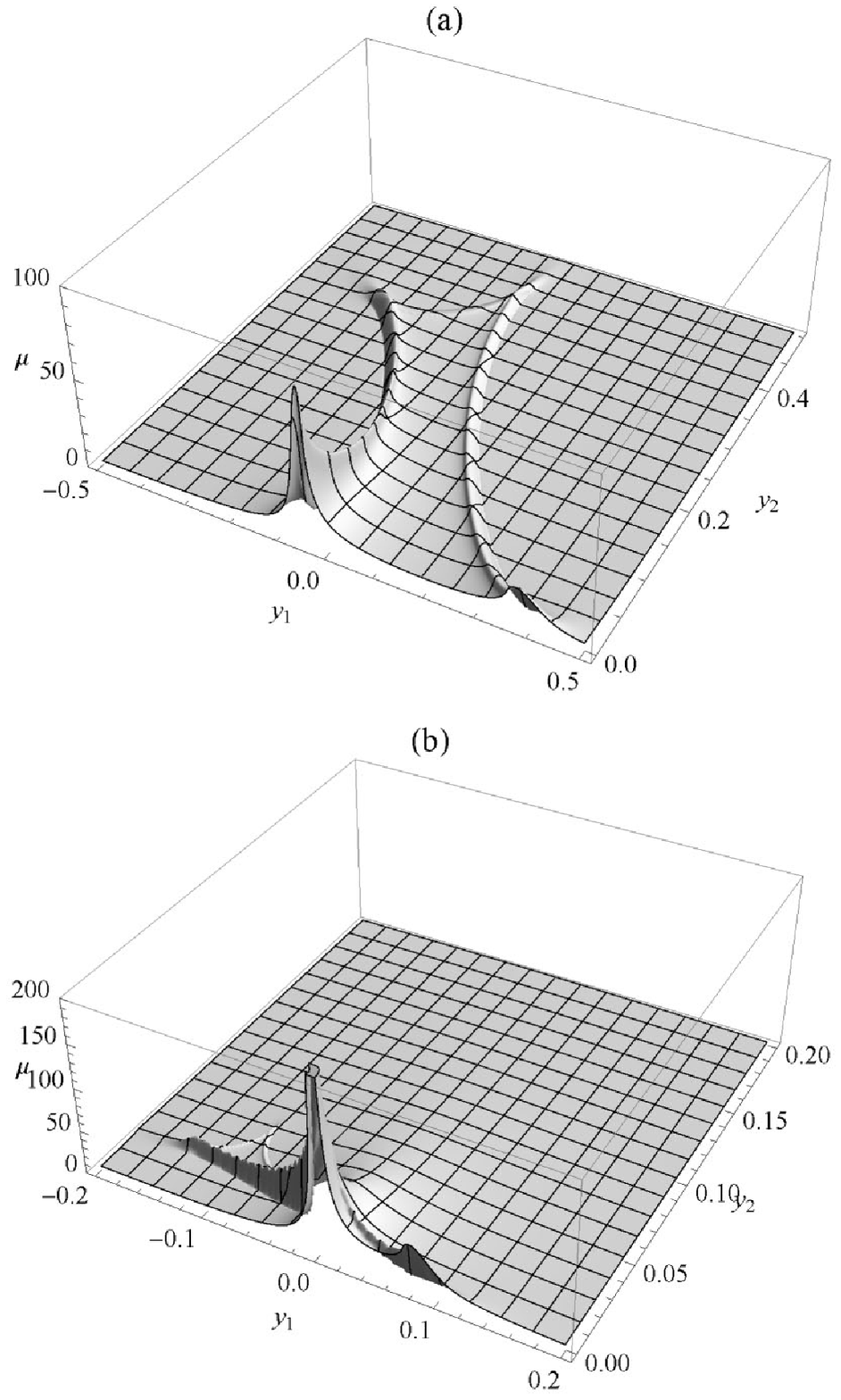}}
 \caption{(a) Magnification map for a binary lens with mass ratio
 $q=0.1$ and separation $d=0.95$; the source radius is
 $\rho_*=0.01$; the step is $\Delta y_{1,2}=0.0025$.
 (b) Magnification map for a binary lens with mass ratio
 $q=0.001$ and separation $d=0.95$; the source radius is $\rho_*=0.001$; the step is $\Delta y_{1,2}=0.001$.}
 \label{Fig mag}
\end{figure}

\section{Numerical examples}

In this section we will consider some explicit examples of
magnification computations with the aim of illustrating the power
of the innovations proposed in the previous sections.

\subsection{Testing the error estimate}

In order to present a test as exhaustive as possible, we calculate
magnification maps with different levels of target accuracy
$\delta \mu$ and evaluate the relative difference. We take maps
calculated at $\delta \mu=10^{-5}$ as reference maps. Maps
calculated at $\delta \mu=10^{-2}$ should not deviate from the
reference maps by more than $10^{-2}$ in order to declare our
error estimate successful. On the other hand, we do not want the
deviation to be too small either, because this would mean that we
are making more calculations than required for matching our target
accuracy. In this subsection we are not considering limb darkening
because we want to focus on the accuracy of the single contour
calculation.

As a first example, we shall consider a binary lens with mass
ratio $q=0.1$ and separation $d=0.95$ (intermediate caustic
topology). Fig. \ref{Fig mag}a shows the reference magnification
map obtained with a source radius $\rho_*=0.01$ and a step size
$\Delta y=0.0025$ on the source plane in both directions. The
caustic is very clearly visible, with a spike in the $\vec y =0$
position, where we get the maximum magnification.

As a second example, we consider a planetary lens with $q=0.001$
and $d=0.95$ (resonant caustic topology). We choose the resonant
caustic topology since it corresponds to the maximum caustic
extension, allowing finer and more stringent tests. The reference
map for $\rho_*=0.001$ and $\Delta y=0.001$ is shown in Fig.
\ref{Fig mag}b. The central spike is much higher in this case,
because the source radius is much smaller (which is necessary for
a better probing of the caustic).

\begin{figure}
\resizebox{10cm}{!}{\includegraphics{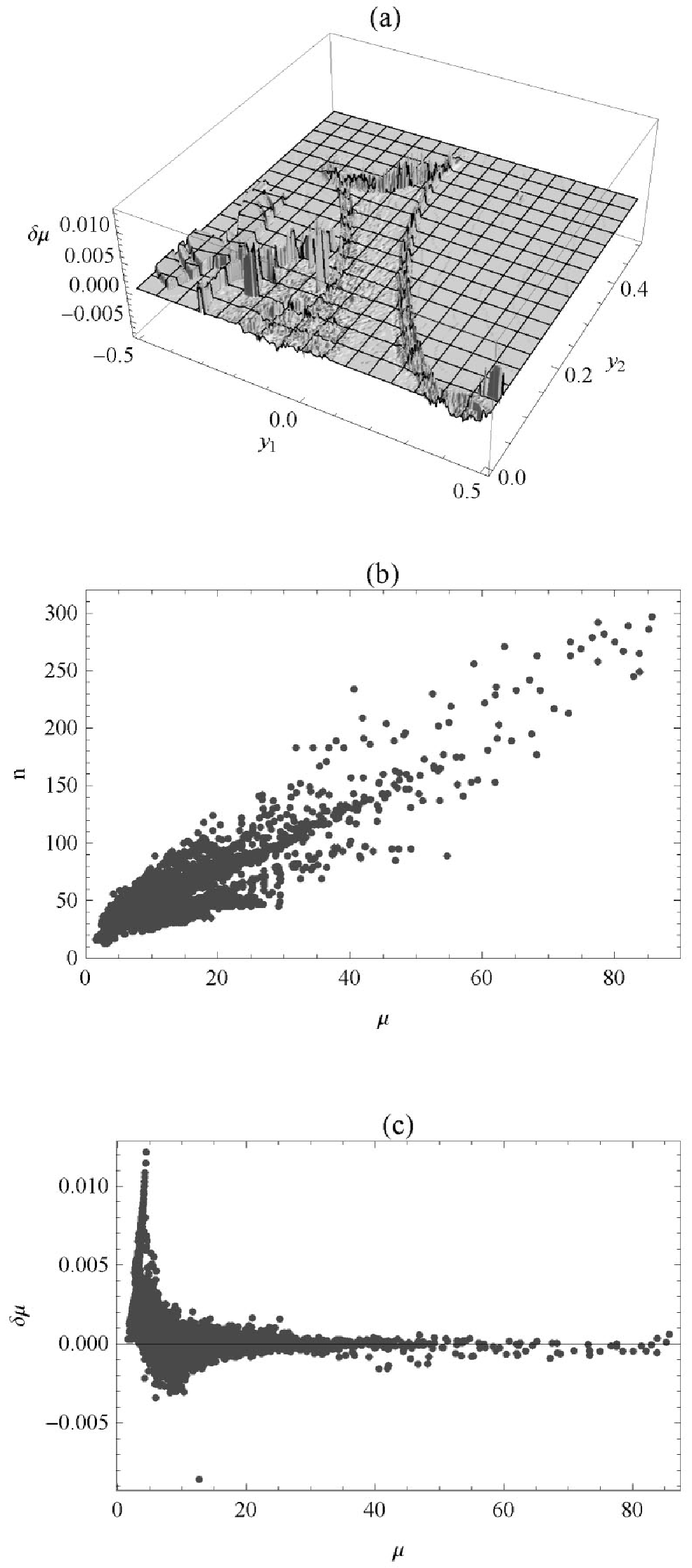}}
 \caption{(a) Map of the errors for a binary lens with mass ratio
 $q=0.1$ and separation $d=0.95$; the source radius is $\rho_*=0.01$;
 the target accuracy is $\delta \mu=10^{-2}$.
 (b) Number of points used in sampling vs magnification.
 (c) Matched accuracy vs magnification. }
 \label{Fig err 1 2 2}
\end{figure}

Now, let us come to the first test. In Fig. \ref{Fig err 1 2 2}a
we show the magnification difference map between a map calculated
at $\delta \mu=10^{-2}$ and the reference map shown in Fig.
\ref{Fig mag}a. We can note that deviations tend to be spatially
correlated far from the caustic, while they are very noisy at
caustic crossings. However, as it is evident from Fig. \ref{Fig
err 1 2 2}c, the target accuracy is fully achieved by all points
in the magnification map. There are just six points with $\delta
\mu>10^{-2}$, with the maximum error being $\delta \mu=0.012$. We
can consider this number of points with slightly exceeding error
acceptable. We can also note that higher magnification points in
the map tend to have smaller errors, with deviations staying one
order of magnitude less than the required accuracy. Without the
parabolic correction and the exit prescription described at the
end of Section \ref{Sec Sampling}, the discrepancy between the
errors of low and high magnification points would be much higher,
so we consider this as a good result of our error estimate
strategy. We can barely see something like a damped oscillatory
behavior of the plot as a function of the magnification. Finally,
in Fig. \ref{Fig err 1 2 2}b we plot the number of sampling points
versus the magnification. The number of sampling points needed for
matching a fixed accuracy $\delta \mu$ grows almost linearly with
magnification, which is what we expect in a Green's theorem
approach.

\begin{figure}
\resizebox{10cm}{!}{\includegraphics{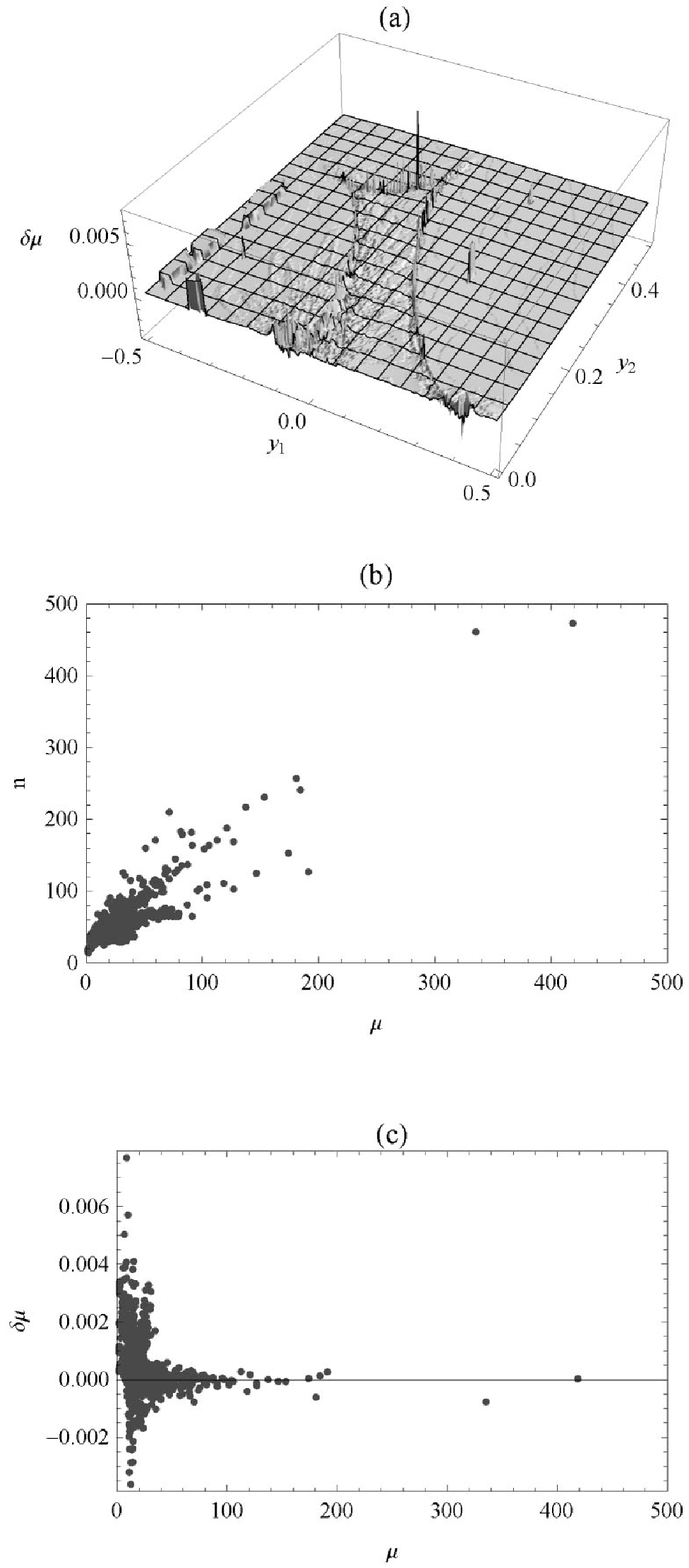}}
 \caption{Same as Fig. \ref{Fig err 1 2 2} with a source radius $\rho_*=0.001$.}
  \label{Fig err 1 3 2}
\end{figure}

Reducing the source size from $\rho_*=0.01$ to $\rho_*=0.001$ has
a slightly beneficial effects on the errors, which however stay at
the correct order of magnitude, as we can see from Fig. \ref{Fig
err 1 3 2}.

\begin{figure}
\resizebox{10cm}{!}{\includegraphics{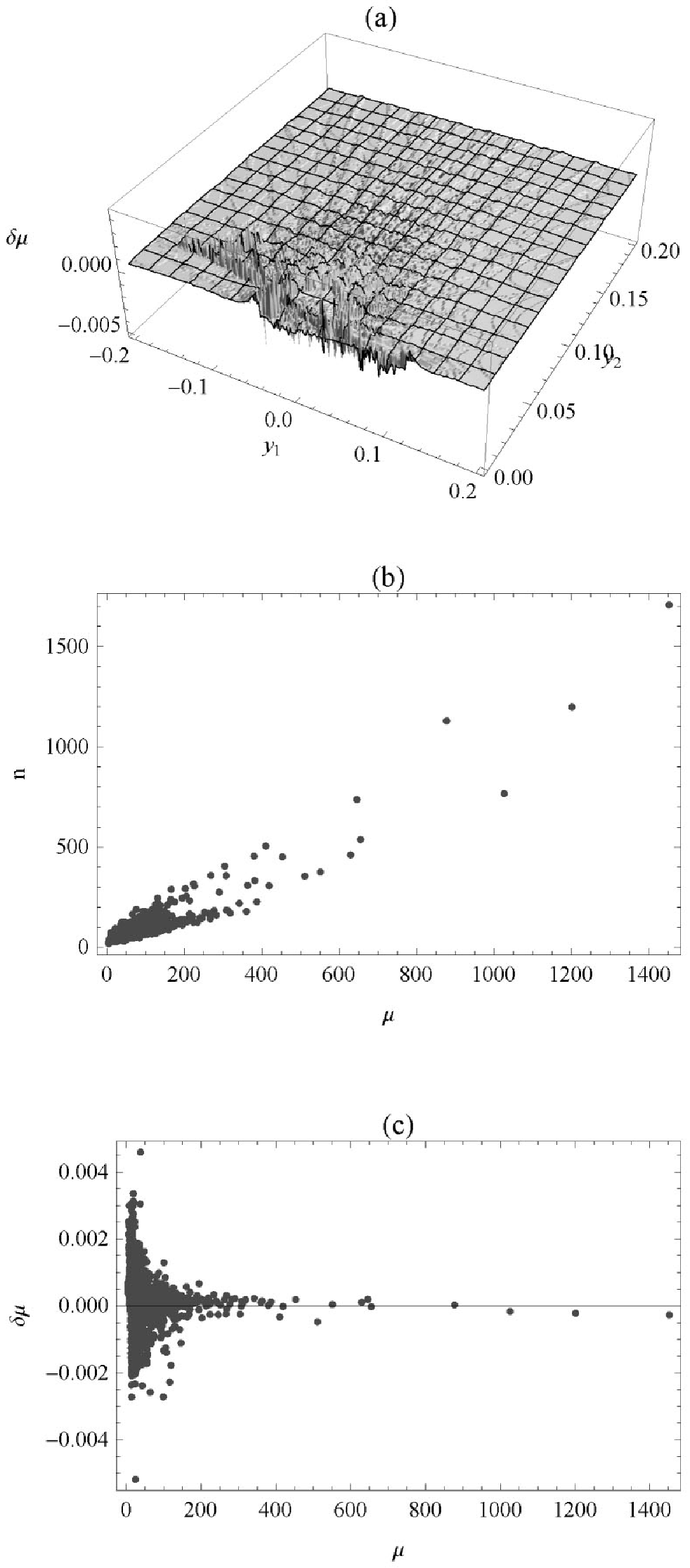}}
 \caption{Same as Fig. \ref{Fig err 1 2 2} with a source radius $\rho_*=0.001$ and a planetary
 mass ratio $q=0.001$.}
 \label{Fig err 3 3 2}
\end{figure}

Coming to the planetary lens, the error map is shown in Fig.
\ref{Fig err 3 3 2}a, where errors appear much more scattered and
less concentrated on the caustic. The errors stay at the correct
order of magnitude and everything seems to be very stable with
respect to the mass ratio.

\begin{figure}
\resizebox{10cm}{!}{\includegraphics{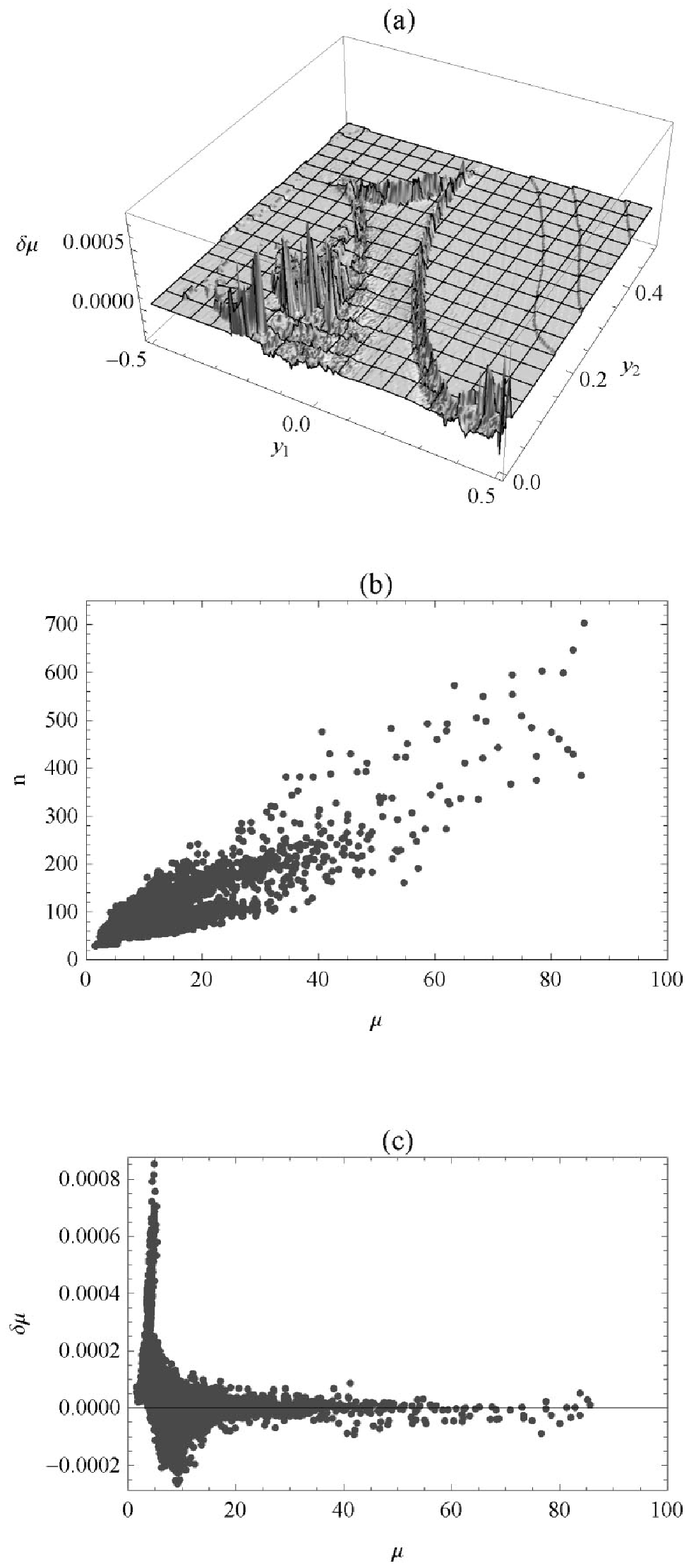}}
 \caption{Same as Fig. \ref{Fig err 1 2 2} with a target accuracy $\delta \mu=10^{-3}$.}
 \label{Fig err 1 2 3}
\end{figure}

Finally, we come back to the mass ratio $q=0.1$ and $\rho_*=0.01$
and try a map with higher target accuracy $\delta \mu=10^{-3}$.
Fig. \ref{Fig err 1 2 3} shows that our sampling strategy and our
error estimate performs in a very successful way at any target
accuracy, with all deviations having the correct order of
magnitude.

\begin{figure}
\resizebox{10cm}{!}{\includegraphics{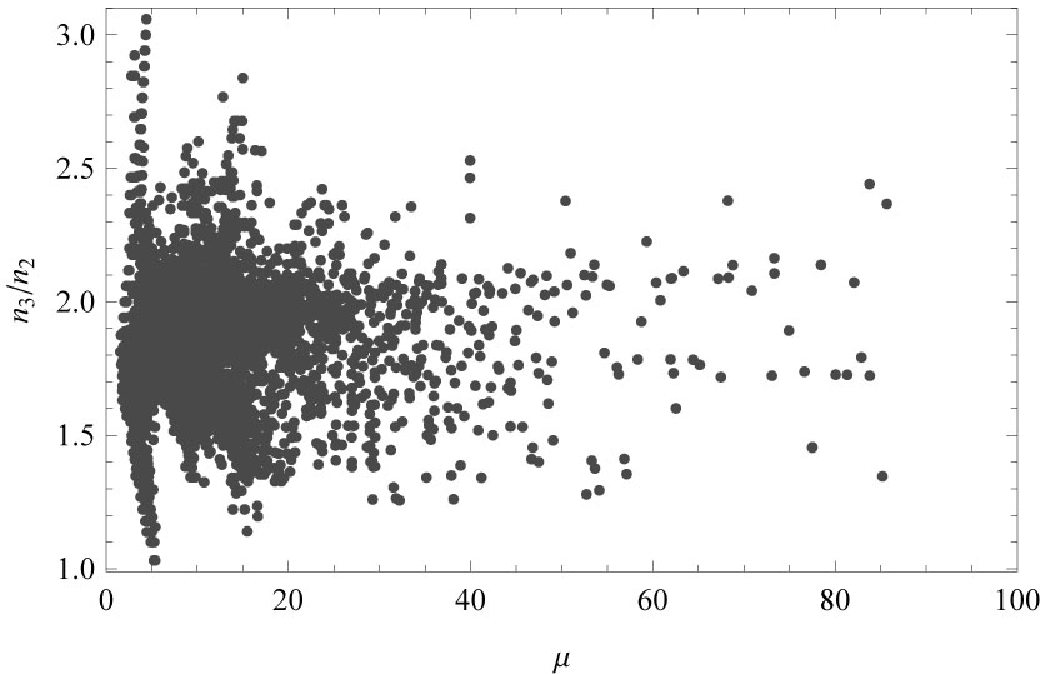}}
 \caption{Ratio of the number of sampling points for an accuracy $\delta \mu=10^{-3}$
 and the number of sampling points with $\delta_\mu=10^{-2}$. The ratio is plotted
 versus the magnification.}
 \label{Fig n3 n2}
\end{figure}

Since the plots of Fig. \ref{Fig err 1 2 2} and \ref{Fig err 1 2
3} only differ for the target accuracy, it is interesting to see
how many sampling points we need to add to go from $\delta
\mu=10^{-2}$ to $\delta \mu=10^{-3}$. In Fig. \ref{Fig n3 n2} we
plot the ratio between the number $n_3$ of sampling points  with
$\delta \mu=10^{-3}$ and the number $n_2$ of sampling points  with
$\delta \mu=10^{-2}$ versus the magnification. Firstly, we see
that the ratio does not depend on the magnification. Secondly, we
can see that the number of sampling points is roughly doubled in
order to increase the accuracy by a factor of 10. More precisely,
the average factor $<n_3/n_2>=1.77$ in our maps. This can be
understood analytically as follows. Considering that the residual
error of the parabolic correction goes as $\Delta\theta^5$ but the
number of sampling points in the interval $[0,2\pi]$ goes as
$\Delta\theta^{-1}$, the accuracy in the magnification goes as
$n^{-4}$. If $n$ is doubled, the accuracy is improved by a factor
16 (it would be just $n^2=4$ without the parabolic correction).
With $<n_3/n_2>=1.77$ we have
$<\delta\mu_2/\delta\mu_3>=<n_3/n_2>^4=9.81$, which is very close
to the ratio of the target accuracies of the two maps.

\subsection{Linear vs Parabolic approximation}

\begin{figure}
\resizebox{10cm}{!}{\includegraphics{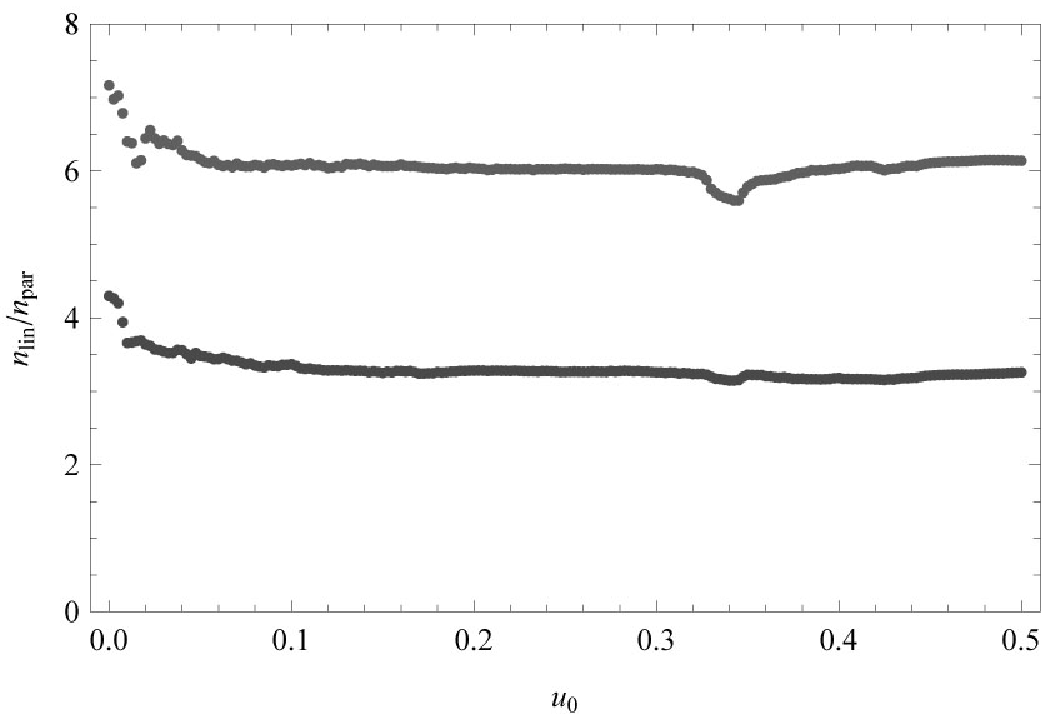}}
 \caption{Speed-up due to the parabolic correction for
 microlensing trajectories parallel to the $y_1$ axis and $u_0$ varying from 0 to
 0.5 (the reference map is in Fig. \ref{Fig mag}a). The upper points are for target accuracy $\delta
 \mu=10^{-3}$ and the lower points are for $\delta \mu=10^{-2}$.}
 \label{Fig nlin npar}
\end{figure}

In the previous subsection we have demonstrated that our error
estimators give us a full control of the accuracy of our
calculations. Now we can go into more detail and try to evaluate
the speed-up due to the parabolic correction.

In order to give an estimate as realistic as possible, we should
consider the calculation of typical microlensing light curves, i.e
we should sum up the time spent for calculating the magnification
along straight lines in the source plane. A good sample of
straight lines is provided by the rows of our magnification maps.
In fact, each row at constant $y_2$ can be considered as a
straight source trajectory parallel to the $y_1$ axis with impact
parameter $u_0=y_2$. Let us denote the number of sampling points
on each row by $n_{par}(u_0)$. This number is proportional to the
total time spent for calculating the full row at $y_2=u_0$.

Similarly, we shall denote the number of sampling points in the
analogous linear calculation (without the parabolic correction) by
$n_{lin}(u_0)$. In order to compare analogous calculations, the
target accuracy $\delta \mu$ must be the same in both cases. As
error estimator in the linear case, we use the parabolic
correction itself, which is already available in our code.

The ratio $n_{lin}/n_{par}$ is thus a measure of the speed-up
obtained by the introduction of the parabolic correction. Fig.
\ref{Fig nlin npar} shows this quantity as a function of $u_0$ for
a target accuracy of $\delta \mu=10^{-3}$ (upper points) and
$\delta \mu=10^{-2}$ (lower points). The number of sampling points
drops by a factor 3.3 in average for $\delta \mu=10^{-2}$ and 6.1
for $\delta \mu=10^{-3}$. The speed-up is even higher for central
events with $u_0 \simeq 0$, in which high-magnification points
have a considerable weight. Such numbers are very encouraging,
since a factor 6 may bring the computational time e.g. for a huge
Markov chain from one weak to a single day.

\subsection{Uniform vs Optimal sampling}

\begin{figure}
\resizebox{10cm}{!}{\includegraphics{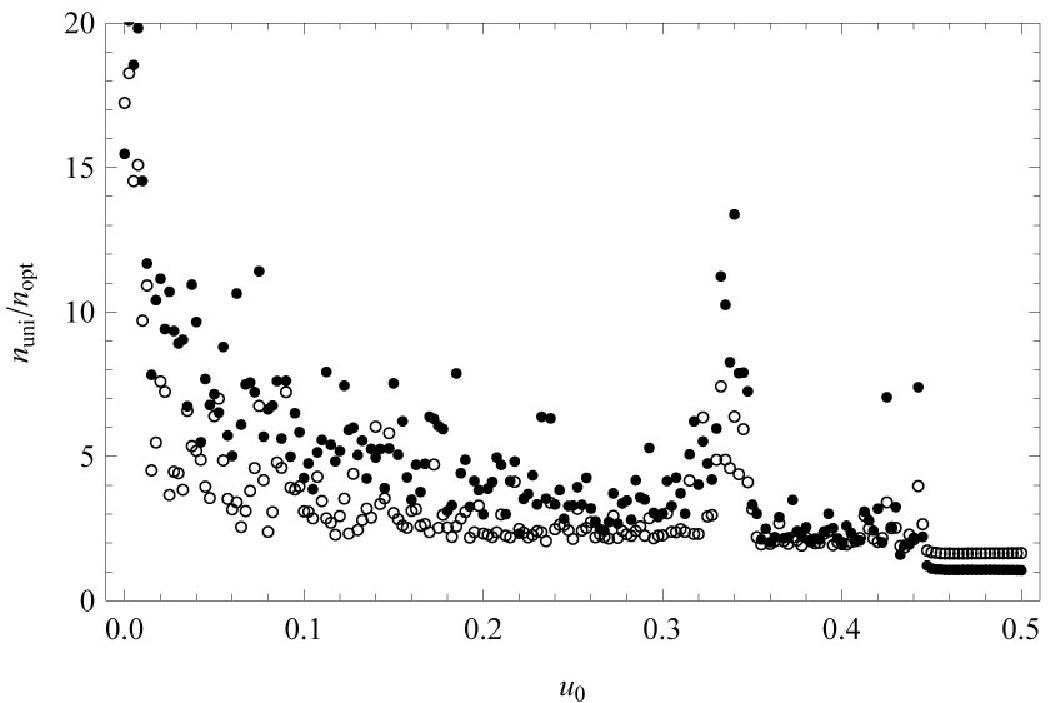}}
 \caption{Speed-up due obtained thanks to optimal sampling for
 microlensing trajectories parallel to the $y_1$ axis and $u_0$ varying from 0 to
 0.5 (the reference map is in Fig. \ref{Fig mag}a). Empty circles are for target accuracy $\delta
 \mu=10^{-2}$; filled circles are for $\delta \mu=10^{-3}$.}
 \label{Fig nuni nopt}
\end{figure}

The next innovation proposed in this paper is the optimal sampling
driven by a reliable estimate of the residual error in each arc
between two sampling points. Indeed, by increasing the sampling
only where really needed (e.g. close to caustic crossings of the
source boundary), we expect to save a considerable amount of
computational time.

In order to evaluate the speed-up, we adopt the same strategy
explained in the previous subsection: we sum up the number of
sampling points for each point in the magnification map at fixed
$y_2$, so that we can have a measure of the time needed to
calculate a microlensing light curve for a source trajectory
parallel to the $y_1$ axis and with $u_0=y_2$. We denote the
number of sampling points with the optimal sampling strategy by
$n_{opt}$ and the number of sampling points with a uniform
sampling by $n_{uni}$. The number of sampling points in the
uniform case is determined by doubling the initial two-points
sampling $\{0,\pi\}$ until the new magnification differs from the
previous one by less than $\delta \mu/2$, where $\delta \mu$ is
the target accuracy. Note that this prescription is sometimes
unsafe, since we can have very small features in the images, which
could be completely missed. However, for the purpose of a gross
estimate of the speed-up, we adopt this prescription for the
uniform sampling, since at most we are just underestimating the
correct speed-up in some points.

In Fig. \ref{Fig nuni nopt} we plot the ratio $n_{uni}/n_{opt}$ as
a function of $u_0$ for trajectories parallel to the $y_1$ axis.
The binary lensing geometry is that described in Fig. \ref{Fig
mag}a. We can see that the speed-up reaches 20 for central
trajectories and then falls down for larger impact parameters.
Indeed, when the source is poorly magnified, there is no need for
optimal sampling. The average speed-up is 3.3 for a target
accuracy $\delta \mu=10^{-2}$ and 4.8 for $\delta \mu=10^{-3}$,
though high magnification events get a much larger benefit from
optimal sampling.

\subsection{Limb Darkening}

\begin{figure}
\resizebox{7cm}{!}{\includegraphics{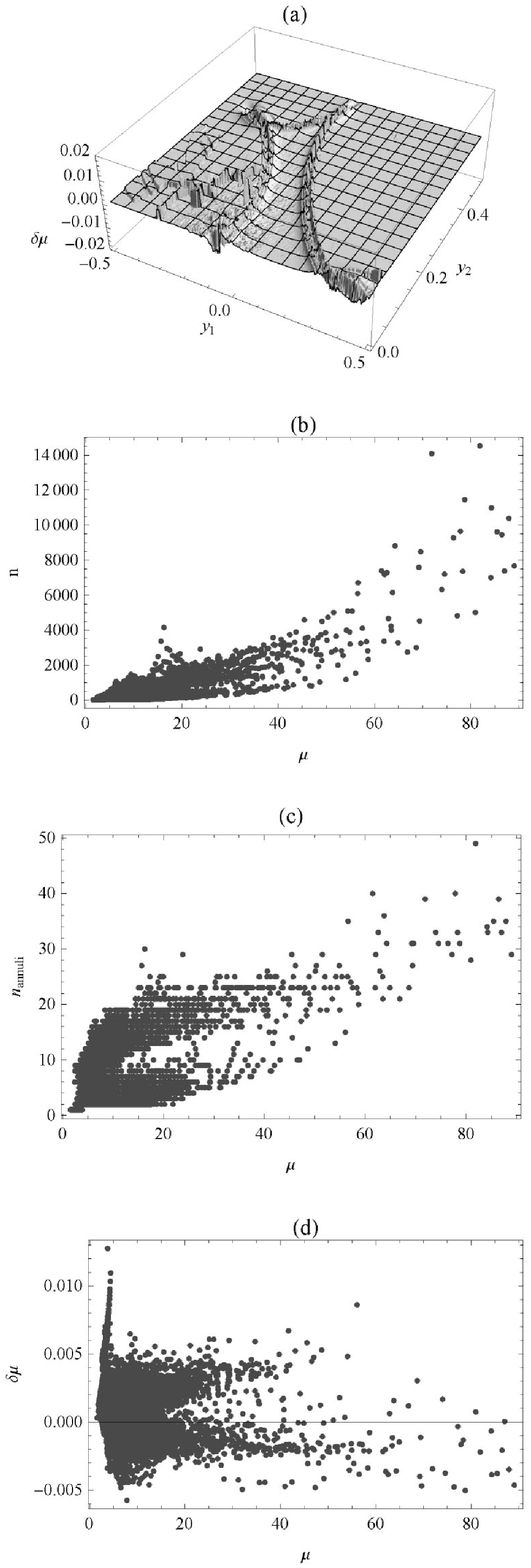}}
 \caption{(a) Map of the errors for a binary lens with mass ratio
 $q=0.1$ and separation $d=0.95$; the source radius is $\rho_*=0.01$ with a linear limb darkening $a=0.51$;
 the target accuracy is $\delta \mu=10^{-2}$.
 (b) Number of points used in sampling vs magnification. (c)
 Number of annuli vs magnification.
 (d) Matched accuracy vs magnification. }
 \label{Fig dark}
\end{figure}

\begin{figure}
\resizebox{10cm}{!}{\includegraphics{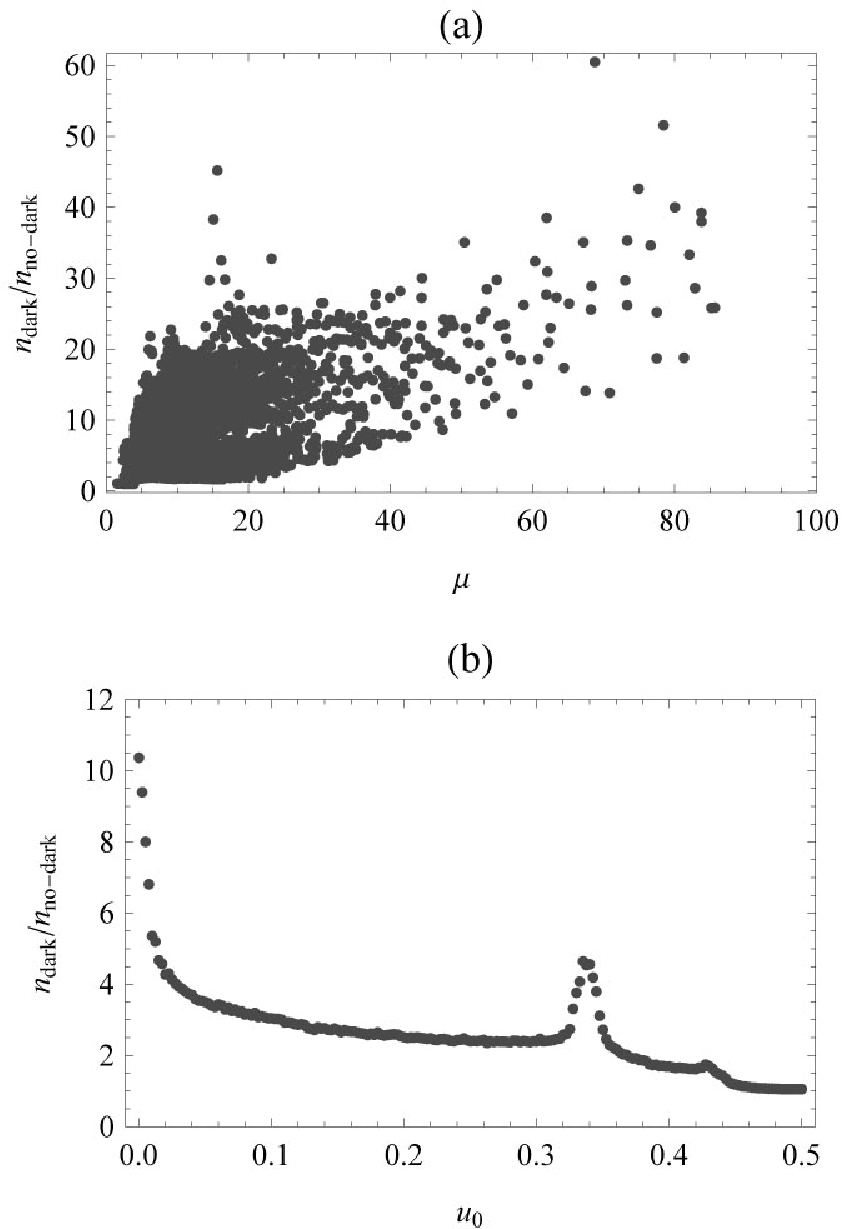}}
 \caption{(a) Ratio between the number of points used with and without the limb darkening treatment.
 (b) Slow-down due to the limb darkening treatment for
 microlensing trajectories parallel to the $y_1$ axis and $u_0$ varying from 0 to
 0.5.}
 \label{Fig dark par}
\end{figure}

As a final test of our code, we consider a linear limb darkened
source with $a=0.51$ and radius $\rho_*=0.01$. We have generated a
reference magnification map with target accuracy $\delta
\mu=10^{-4}$. We do not show it because it looks very similar to
Fig. \ref{Fig mag}a, except for the height of the magnification
peaks. After that, we have generated a test magnification map with
target accuracy $\delta \mu=10^{-2}$. In both cases we have used
the error estimators introduced in section \ref{Sec Limb} and the
optimal source sampling strategy described there.

The difference between the test and the reference map is shown in
Fig. \ref{Fig dark}a. We can see that errors are kept well under
control, in particular at caustic crossings, which represent the
most crucial tests for limb darkening. Of course, approximating
limb darkening by the contour method requires a large number of
points. In Fig. \ref{Fig dark}b we have a plot of the number of
sampling points (adding those in all annuli) vs the magnification.
We can see that the number of points grows faster than linearly
with magnification at large $\mu$. The number of annuli (Fig.
\ref{Fig dark}c) grows rapidly from 1 to 20 at low magnifications
and then stays more or less constant up to very high
magnifications. Finally, Fig. \ref{Fig dark}d shows that the
target accuracy has been achieved by all points save for three
with $\delta \mu =0.011$.

The bottom line of our limb darkening code is the slow-down plot
shown in Fig. \ref{Fig dark par}. We compare the number of points
required for a calculation of limb darkening by summing up the
number of sampling points in all annuli. We denote this number by
$n_{dark}$. We compare this number with the number of sampling
points in the uniform source case ($n_{no-dark}$). In panel (a) we
show the ratio $n_{dark}/n_{no-dark}$ vs the magnification.
Furthermore, as in previous subsections, we sum up the total
number of sampling points along each row at fixed $y_2$ in our
magnification map. This is an indication of the total time spent
for calculating a full microlensing trajectory parallel to the
$y_1$ axis with $u_0=y_2$. The ratio of the number of points
including the limb darkening treatment $n_{dark}$ and the number
of points calculated for a uniform brightness disk $n_{no-dark}$
therefore represents the slow-down factor of our code for the
inclusion of the limb darkening treatment. From Fig. \ref{Fig dark
par}b we see that the average slow-down factor is 2.64, but
central trajectories may be slowed down to a factor of 11.
Interestingly, we also have two peaks at impact parameters
$u_0=0.34$ and $u_0=0.42$, which correspond to trajectories
including cusp crossings.

\section{Conclusions}

In this paper we have presented four innovations for codes
attempting microlensing calculations based on contour integration.
In this class of codes the contours of the images are obtained by
inversion of the lens map at the source boundary; the area of the
images is then obtained by a simple contour integration rather
than by a surface integration.

We have introduced a parabolic third order correction in the
evaluation of Green's line integral, which leaves a residual of
the fifth order in the interval size $\Delta \theta$. The speed-up
with respect to the classical trapezium approximation is around 4
for a target accuracy in the magnification of $\delta
\mu=10^{-2}$.

We have introduced accurate error estimators, whose reliability
has been shown by comparing several magnification maps obtained at
different levels of target accuracy.

Thanks to these error estimators, we have proposed a new optimal
sampling strategy driven by the error estimates in each sampling
interval. With respect to a uniform sampling we get a speed-up
ranging from 3 to 20 at $\delta \mu=10^{-2}$ depending on how
large is the magnification experienced by the source in its
microlensing trajectory.

Finally, we have faced the problem of limb darkening, which is the
hardest obstacle for codes based on contour integration. Also in
this case we have introduced error estimators and an optimal
sampling strategy with the aim of minimizing calculation keeping
full control of the accuracy. As a result, we have a very reliable
limb darkening approximation to any desired accuracy. At $\delta
\mu=10^{-2}$ the slow-down with respect to a uniform brightness
disk ranges from 2 to 11.

Summing up, the speed-up gained by the optimal sampling is
sufficient to compensate the slow-down due to the migration from
uniform brightness disks to realistic limb darkened sources. In
addition, the parabolic correction guarantees a net speed-up with
respect to traditional linear Green's theorem codes. Finally, the
full control of the errors is an invaluable help in order to avoid
redundant calculations and concentrate the efforts where it is
really needed. Its impact with respect to traditional codes with
fixed sampling is definitely huge and difficult to quantify.

All these innovations should put codes based on contour
integration in the front line for binary and planetary
microlensing events modelling.

\acknowledgments

I wish to thank Martin Dominik for his precious suggestions and
comments on the manuscript. I also thank the MiNDSTEp consortium
for giving me the opportunity to confront my code with real
microlensing data. I acknowledge support by PRIN (Prot.
2008NR3EBK\_002), and research funds of Salerno University.

\end{document}